\documentclass{article}
\usepackage{icmc2019template}
\usepackage{amsmath}
\DeclareMathOperator*{\argmax}{arg\,max}
\usepackage{commath}
\usepackage{times}
\usepackage{ifpdf}
\usepackage{soul}
\usepackage[english]{babel}

\def\papertitle{An Attentional Neural Network Architecture for Folk Song Classification}
\def\firstauthor{Aitor Arronte Alvarez}
\def\secondauthor{Francisco G\'omez-Mart\'in}
\def\thirdauthor{Third Author}

\newif\ifpdf
\ifx\pdfoutput\relax
\else
   \ifcase\pdfoutput
      \pdffalse
   \else
      \pdftrue
  \fi
\fi

\ifpdf 
  \usepackage[pdftex,
    pdftitle={\papertitle},
    pdfauthor={\firstauthor, \secondauthor, \thirdauthor},
    bookmarksnumbered, 
    pdfstartview=XYZ 
   ]{hyperref}

  \usepackage[pdftex]{graphicx}
  \graphicspath{{./figures/}}
  \DeclareGraphicsExtensions{.pdf,.jpeg,.png}

  \usepackage[figure,table]{hypcap}

\else 
  \usepackage[dvips,
    bookmarksnumbered, 
    pdfstartview=XYZ 
  ]{hyperref}  

  \usepackage[dvips]{epsfig,graphicx}
  \graphicspath{{./figures/}}
  \DeclareGraphicsExtensions{.eps}

  \usepackage[figure,table]{hypcap}
\fi

\hypersetup{
    colorlinks,%
    citecolor=black,%
    filecolor=black,%
    linkcolor=black,%
    urlcolor=black
}

\title{\papertitle}

\twoauthors
{1.5in}
{\firstauthor} {Center for Language and Technology\\ University of Hawaii at Manoa\\  %
{\tt \href{mailto:arronte@hawaii.edu}{arronte@hawaii.edu}}}
{\secondauthor} {Applied Mathematics Department\\ Technical University of Madrid\\  %
{\tt \href{mailto:fmartin@etsisi.upm.es}{fmartin@etsisi.upm.es}}}

\begin{document}
\capstartfalse
\maketitle
\capstarttrue
\begin{abstract}
In this paper we present an attentional neural network for folk song classification. We introduce the concept of musical motif embedding, and show how using melodic local context we are able to model monophonic folk song motifs using the skipgram version of the word2vec algorithm. We use the motif embeddings to represent folk songs from Germany, China, and Sweden, and classify them using an attentional neural network that is able to discern relevant motifs in a song. The results show how the network obtains state of the art accuracy in a completely unsupervised manner, and how motif embeddings produce high quality motif representations from folk songs. We conjecture on the advantages of this type of representation in large symbolic music corpora, and how it can be helpful in the musicological analysis of folk song collections from different cultures and geographical areas.
\end{abstract}

\section{Introduction}\label{sec:introduction}
The increasing availability of digital music corpora and the growing interest in empirical approaches and methods in musicology has brought new challenges and opportunities for Musical Information Retrieval (MIR). Large symbolic cross-cultural music corpora demand new tools that can extract relevant information in an automated manner. In this paper we are interested in researching the possibilities of using vector representations of musical patterns based on their context. Having a vector representation of a musical entity such as a motif, will allow for the direct comparison of patterns and contexts using the cosine similarity measure. This approach pretends to facilitate the musicological analysis by using machine learning vector embedding techniques to extract similar patterns and their contexts from large collections of symbolic music databases.

Vector representations of words, or word embeddings, have had a great success in Natural Language Processing (NLP) tasks \cite{rumelhart1986learning}. Based on the idea that words that are semantically similar to each other are represented closer in a continous vector space, the word2vec algorithm has shown the ability to represent high-quality word embeddings from large text corpora \cite{mikolov2013efficient,mikolov2013distributed, goldberg2014word2vec}. NLP methods have been adopted and adapted in MIR contexts  \cite{conklin1995multiple}, \cite{boulanger2012modeling}, \cite{de2017large}. Word2vec was used to model musical contexts in western classical music works \cite{herremans2017modeling}, and for chord recommendations \cite{huang2016chordripple}. In this paper we deal with a more limited data context, monophonic folk songs. 

Our goal is to adopt the skip-gram version of the word2vec model for the distributional representation of motifs. Several melodic features such as contour, grouping, and small size motifs seem to be part of the so called ‘Statistical Music Universals’ \cite{nettl2000ethnomusicologist}, \cite{savage2015statistical}. This sequential processing of melodic units may be related to the human capacity to group and comprehend motifs as units within a melodic context. Our hypothesis is that these units may relate to each other in a melody in similar ways as words do in sentences. If that is the case, the word2vec algorithm should be able to represent motifs from folk songs. The motif embeddings will be used as the input in a classification task using an attentional neural network architecture.

Deep learning methods for text classification such as convolutional neural networks \cite{kim2014convolutional}, and recurrent neural
networks based on long short-term memory (LSTM) \cite{hochreiter1997long} have proven to be very effective. Encoder-decoder methods from the Machine Translation literature, where an encoder neural network reads and encodes a sentence into a fixed-length vector, and an decoder outputs a translation of the sentence by decoding the initial representation. One of the shortcoming of this approach is the fact that sentences are encoded as a fixed-length vector, and in a corpus where sentences greatly vary in size, the performance of this method deteriorates quickly \cite{cho2014learning}.  An attentional mechanism that searches for a set of positions in an encoded sentence where the most relevant information is kept was presented to overcome this limitation \cite{bahdanau2014neural}. The relevant information is preserved in a context vector, so a target word based on this vector can be predicted. We use this so called 'attention' mechanism, to search for motifs that are more relevant than others in a song based on a melodic context.

The remaining of this paper is organized as follows: in section \ref{sec:w2vec} we introduce in formal terms the word2vec model, present how the data is encoded, and show based on ad-hoc queries the quality of the motif embeddings. In section \ref{sec:attention} we present the attentional neural network for classifying folk song based on the motifs obtained with the word2vec algorithm. Section \ref{sec:experiment} details the data used and the experiments, presenting the results in \ref{sec:results}. We conclude in section \ref{sec:conclusions} by highlighting the potential use of this type of representation and classification method in the analysis of large corpora from diverse cultures and geographical areas.

\section{Motif embeddings}\label{sec:w2vec}
\subsection{Word2vec algorithm}
\label{ssec:w2vmodel}
In the skip-gram version of the word2vec model, the goal is to find word embeddings that can predict the surrounding words of a target word in a sentence or document \cite{mikolov2013distributed}. Formally, we can define the model in the following terms: given a corpus $W$ of words $w$ and contexts $c$, the network tries to predict the surrounding words of a target in a context. The objective of the skip-gram is to set the parameter $\theta$ in $p(c \mid w; \theta)$ that maximizes the corpus probability:
\begin{equation}
\argmax_\theta \prod_{w  \in W} \Bigg[  \prod_{c  \in C}  p(c \mid w; \theta) \Bigg]
\end{equation}

where  $p(c \mid w; \theta)$ is calculated by the softmax function:

\begin{equation}
p(c \mid w; \theta)= \frac{e^{v_c \cdot v_w}} {\sum_{c\sp{\prime} \in C} e^{v_{c\sp{\prime} } \cdot v_w}}
\end{equation}

where \(v_c\) and \( v_w \in R^d \) are vector representations of \textit{v} and \textit{c}, and \textit{C} is the set of all possible contexts. The set of parameters \(\theta\) is composed of \( v_{c_i} \), \( v_{w_i} \) for \( w\in W\).

Since the term \( p(w; \theta) \) involves a summation over all possible contexts \( c \sp{\prime} \) it becomes computationally very intensive, and it is normally replaced with negative sampling \cite{mikolov2013distributed}. In this article we use this sampling technique.

We use the cosine similarity measure to determine the relatedness of two embeddings. We define the metric for a pair of words \( w_1 \) and \( w_2 \) as \cite{schnabel2015evaluation} :

\begin{equation}
cos(w_1, w_2)=\frac{\overrightarrow{w_1} \cdot \overrightarrow{w_2}}{\norm{\overrightarrow{w_1}} \norm{\overrightarrow{w_2} }} 
\end{equation}

for all similarity computations in the embedding space, where $\overrightarrow{w}$ is a real-valued vector embedding of word $w$.

\subsection{Motif representation}\label{ssec:representation}

Melodic similarity and classification methods rely strongly in the manner in which music is represented \cite{toiviainen2002computational}. Our goal is to extract motifs from folk songs based on melodic context using the word2vec algorithm. We understand context as the sequential organization of melodic units that establish statistically relevant relationships with one another in a melodic segment. We represent, separately, rhythmic and intervallic motifs using strings.

We codify all the intervals for each song using Music21 \cite{cuthbert2010music21} chromatic step values, and encode interval direction using Boolean values (\texttt{1} for ascending and \texttt{0} for descending). For instance, the string \texttt{21} represents an ascending major second, and the string \texttt{30} a descending minor third. Repeated notes are encoded as \texttt{00}. We represent rhythmic units following a similar approach. We codify with Boolean values whether the value is a rest (\texttt{0}) or a note (\texttt{1}), and the upbeat (\texttt{0}) or downbeat (\texttt{1}). Duration representations are based on the unit of a quarter note. We separate with a dash each of these features. The string \texttt{1-1-0.5} represents then, an eighth note that falls on a downbeat.

The next step to obtain motif embeddings is to discover motifs as multi-words, or prototype words \cite{mikolov2013distributed}. The motivation behind this idea is that, since we are working with the smallest intervallic and rhythmic units, when grouped with each other based on their frequency of occurrence within a corpus, we will obtain statistically relevant motifs. A multi-word is then a concatenation of two or more intervals or durations that are found in a melody adjacent to each other. For example, an intervallic multi-word of size 3 \texttt{30\_00\_21} represents a descending minor third, followed by a repeated note, and by an ascending major second. And the rhythmic multi-word of size 2 \texttt{1-0-0.25\_1-1-0.5} represents a sixteenth note on an upbeat, followed by an eighth note on a downbeat.

Once the vocabulary of multi-words is created, we codify songs using this representation method and apply the skip-gram version of word2vec to obtain vector representations of all motifs in a corpus.

\section{Attentional Neural Network Architecture}\label{sec:attention}

We present a neural network architecture based on the attentional mechanism described in \cite{bahdanau2014neural}, which uses context to determine the importance of a word in a given sentence. We will use this mechanism to search for motifs that are more relevant than others in a song for the correct classification of melodies given their geographical and national collection.

The architecture that we present is composed of a motif encoder, a motif attention layer, and an output single layer Multilayer Perceptron (MLP) to perform the classification task. 

The motif encoder takes an input sequence $X$ and reads the sequence as a vector representation $X=(x_1, \cdots, x_T)$ into a vector $c_i$ where $T$ is the length of the input sequence. Instead of following the order of the sequence from start to finish, in this approach the encoder annotates not only the preceding words of a target word, but also the following words using a bidirectional GRU (BGRU) \cite{bahdanau2014neural}. A BGRU is composed of a forward and backward GRU, where the forward GRU $\overrightarrow{f}$ reads the input as it is ordered and estimates the forward hidden states $\overrightarrow{h_1}, \cdots , \overrightarrow{h_T}$. The backward GRU $\overleftarrow{f}$ obtains a backward representation of the hidden states by processing the sequence in reverse order. An annotation $h_j$ is obtained by concatenating the forward $\overrightarrow{h_j}$ and backward  states $\overleftarrow{h_j}$, which represents the motifs around a target motif $w_j$ in a song or melody.

The motif attention layer computes the following:

\begin{equation}
c_i= \sum_{j=1}^{T}\alpha_{ij}h_j.
\end{equation}
The weight $\alpha_{ij}$ is computed as: 
\begin{align}
\alpha_{ij}= \frac{exp(e_{ij})}{\sum_{k=1}^{T} exp(e_{ik})},
\end{align}
where
\begin{align*}
e_{ij}=a(s_{i-1}, h_j)
\end{align*}

is the alignment of the output at position $i$ that matches the input at position $j$, and $s_{i-1}$ is the previous hidden state. The function $a$ is a MLP trained jointly with the rest of the components of this architecture. The vector $c_i$ summarizes all the information of motifs in a song. We use the negative log likelihood of the correct labels to determine song classification:

\begin{equation}
L= -\sum_{d}\log p(d_l),
\end{equation}
where $l$ is the label of a song $d$.


\section{Data and Experiments}\label{sec:experiment}
\subsection{Data}\label{ssec:data}

To test the proposed architecture we use the Chinese and German corpora from the Essen folksong collection \cite{schaffrath1995essen}, and a collection of Swedish folk songs \cite{swedfolksong}. The corpus with the German and Chinese collections is composed of 4923 melodies (2682 from the German and 2241 from the Chinese), and 489 from the Swedish collection.

We codify all melodies following the procedure described in subsection \ref{ssec:representation}, and obtain two corpora: a corpus of intervallic and a corpus of rhythmical motifs.

\subsection{Experiment 1}\label{ssec:exp1}

We first test our architecture with data from only the Chinese and German collections, replicating the experiment presented in a previous work \cite{cuthbert2011feature}, where the authors classify songs as being from either China or Central Europe (mostly Germany) using a hand-crafted set of 174 features. The main difference with our approach is that feature extraction and classification is performed in a completely unsupervised manner. The musical features in our case would be the vector representations of the motifs for each song extracted automatically using word2vec, and the unsupervised classification is performed by the proposed network architecture. The number of songs to classify in this experiment is much larger (4923) compared to the 1943 songs used in the previous work \cite{cuthbert2011feature}. We split the dataset in training (75\text{\%} of the total data, which results in 3692 melodies), and test set (1231 melodies).

We compare the classification accuracy of our model with two baselines: paragraph vectors (doc2vec) \cite{le2014distributed}, and an average of the vectors obtained with the word2vec algorithm. We obtain song representations for each one of these two models and perform a binary classification task using a linear SVM classifier. 

We implement the proposed architecture using the Keras framework \cite{chollet2015keras}.

\subsection{Experiment 2}\label{ssec:exp2}

The second experiment uses the 3 folk song collections in their entirety (5412 songs). We analyze performance accuracy by class, and test whether our model is able to predict correctly using an unbalance corpus, where two of the classes belong to two closely related European folk song collections.

\begin{table}
\begin{center}
\begin{tabular}{ |l||l|l|}
 \hline
 \multicolumn{3}{|c|}{Intervallic Motif Embeddings} \\
 \hline
 Motif 1& Motif 2&Cosine distance\\
 \hline
\texttt{20\_00} & \texttt{00\_20}  & 0.9787\\

\texttt{10\_20}   & \texttt{20\_10}   & 0.9771\\

\texttt{21\_20\_20}   & \texttt{20\_21\_20}   &0.996\\

\texttt{40\_21\_21}   & \texttt{21\_21\_40}  &0.998\\

 \hline
\end{tabular}
\end{center}
\caption{\textit{Comparison of motif pairs based on cosine similarity.}} \label{tab:motifs}
\end{table}

\section{Results}\label{sec:results}

\subsection{Motif embeddings}\label{ssec:motem}

Table \ref{tab:motifs} shows the most similar pairs of motifs obtained from the data described in \ref{ssec:data} using  the word2vec algorithm and ad hoc queries. Cosine similarity measures show how melodic context alone can be used to model high quality motif embeddings. All intervallic motifs in the table have the same intervals, either in reverse order or by altering the descending direction with ascending or viceversa. For instance, motid   \texttt{21\_20\_20} is composed of an ascending major second and two descending major seconds. The most similar motif, with cosine similarity of 0.996, is a permutation of the first and second interval \texttt{20\_21\_20}.

\subsection{Experiments}\label{ssec:expres}

We use the motif embeddings for multiwords $mw$ of sizes 2 and 3, for melodic, and rhythmical features with a vector dimension embedding of 150 as the input for our architecture. The motif encoder and attention layer in the architecture have 200 hidden nodes. We use a stochastic gradient descent algorithm (SGD) to train the model with mini-batches of size 10 (the most optimal size found).

Table \ref{tab:comparison} shows classification precision scores for experiment 1. We compare the classification results for intervallic, and rhythmical representations of the songs using the proposed architecture with the two base models. The attention network outperforms both baselines in all types of representations and for $mw$ of size 2 and 3. Smaller interval motifs seem to be the best representation in terms of classification precision. We should note the good quality of the results in general not only for the proposed archtecture, but also for the doc2vec model, which highlights the quality of the motif representation and its impact on classification accuracy.

The results from the proposed architecture obtain similar results to the best model (96\text{\%}  accuracy) from the previous work \cite{cuthbert2011feature}, with the difference that we do not use hand-crafted features, and our corpus is much larger. Our classification task is reduced to Chinese and German folk songs, instead of Chinese and European collections as a whole.

The second experiment tests the architecture in a more nuanced classification task. We include the Swedish folk song collection to the corpus, taking into consideration that European folk music is considered by many as a single corpus of musical style \cite{nettl1990folk}. The results in Table \ref{tab:fullclass} show that even with an unbalance corpus and with 2 of the 3 classes coming from a similar collection, we obtain results that are only 2.4\text{\%} less accurate than the best result in experiment 1. Table \ref{tab:correctclass} shows precision scores by class. We can see how the most differentiated collection, the Chinese, has better classification accuracy.

\begin{table}
\begin{center}
\begin{tabular}{ |l||l|l|l|}
 \hline
 \multicolumn{4}{|c|}{Model Comparison} \\
 \hline
 Repres. type& mw size&Model&Precision\\
 \hline
 Intervallic   & 2   &Attention network & \textbf{0.9458}\\
 Intervallic   & 2   &Doc2vec &0.9142\\
 Intervallic   & 2   &Average word2vec &0.8947\\
 
Rhythmic   & 2   &Attention network & \textbf{0.9279}\\
Rhythmic  & 2   &Doc2vec &0.8967\\
Rhythmic   & 2   &Average word2vec &0.8189\\

Intervallic   & 3   &Attention network & \textbf{0.9341}\\
Intervallic   & 3   &Doc2vec &0.9243\\
Intervallic   & 3   &Average word2vec &0.9019\\
 
Rhythmic   & 3   &Attention network & \textbf{0.9116} \\
Rhythmic  & 3   &Doc2vec &0.8578\\
Rhythmic   & 3   &Average word2vec &0.8009\\
 \hline
\end{tabular}
\end{center}
\caption{\textit{Binary classification precision scores for each model.}} \label{tab:comparison}
\end{table}

\section{Conclusions}\label{sec:conclusions}

In this article we present an attention based neural architecture for folk song classification. We use the skip-gram version of word2vec to learn rich representations of monophonic folk song motifs from Chinese, German, and Swedish collections. The architecture obtains state of the art results in a completely unsupervised manner, and is  able to classify closely related folk song collections with a high degree of accuracy. Since the method does not require human supervision or exlicit expert knowledge, it can be used for the analysis of large collections of symbolic musical data.

In this research, we also show how motif embeddings capture melodic relations based on local context, and how the use of this type of representation, motif embeddings, can be learned from a large corpus. The cosine similarity can be then used to find variations of the same motifs. This representation can be very useful for the musicological study of folksong variation using small melodic units such as motifs. It also shows, how word2vec is able to capture and model complex melodic features.

Future research should explore the hierarchical organization of songs based not only on motifs, but also on musical phrases. Recent research in NLP has shown how hierarchical representation of text documents improves document classification \cite{yang2016hierarchical}.

\begin{table}
\begin{center}
\begin{tabular}{ |l||l|l|}
 \hline
 \multicolumn{3}{|c|}{Attention Network Results} \\
 \hline
 Repres. type& mw size&Precision score\\
 \hline
 Intervallic   & 2   & 0.922\\

 Rhythmic   & 2   & 0.889\\

Intervallic   & 3   &0.91\\

Rhythmic   & 3   &0.878\\

 \hline
\end{tabular}
\end{center}
\caption{\textit{Results for the Chinese, German, and Swedish collections.}} \label{tab:fullclass}
\end{table}

\begin{table}
\begin{center}
\begin{tabular}{ |l||l||l|}
 \hline
 \multicolumn{2}{|c|}{Precision by Class} \\
 \hline
 Collection&Score\\
 \hline
German   & 0.866\\

Chinese  & 0.957\\

Swedish   &0.921\\

 \hline
\end{tabular}
\end{center}
\caption{\textit{Correct classification by label.}} \label{tab:correctclass}
\end{table}

\bibliography{icmc2019template}

\end{document}